\documentclass[useAMS,usenatbib,referee]{biom2arxiv}

\usepackage{graphicx}
\usepackage{amsmath,amsfonts,amssymb}
\usepackage{enumerate} 
\usepackage[hidelinks]{hyperref} 
\usepackage{natbib} 

\DeclareMathOperator{\Eval}{\mathbb{E}}
\DeclareMathOperator{\OVE}{OE} 
\DeclareMathOperator{\SE}{SE} 
 
\DeclareMathOperator{\OVETV}{\OVE_{B}}
\DeclareMathOperator{\muTV}{\mu_{B}}
\DeclareMathOperator{\omegaTV}{\omega_B}
\DeclareMathOperator{\hmuTV}{\widehat{\mu}_B}
\DeclareMathOperator{\hOETV}{\widehat{\OVE}_B}

\DeclareMathOperator{\mean}{mean}

\DeclareMathOperator*{\argmax}{arg\,min}
 
\DeclareMathOperator{\expit}{\mathcal{L}^{-1}} 
\DeclareMathOperator{\Var}{Var} 
 
\usepackage{filecontents}
 
\title{Causal Inference from Observational Studies with Clustered Interference} 
 
\author[Barkley et al.]{
	\hspace{-0.42 in} 
	 Brian G. Barkley$^{1}$, 
		Michael G. Hudgens$^{1,*}$\email{mhudgens@bios.unc.edu},
		John D. Clemens$^{2}$,  
		Mohammad Ali$^{3}$, and
		Michael E. Emch$^{4}$
		 \\
			\hspace{-0.5 in}	$^{1}$Department of Biostatistics, University of North Carolina, Chapel Hill, North Carolina, U.S.A.\\
			\hspace{-0.5 in}	$^{2}$Department of Epidemiology, University of California, Los Angeles, California, U.S.A.\\		
			\hspace{-0.5 in}	$^{3}$Department of International Health, Johns Hopkins University, Baltimore, Maryland, U.S.A.\\	
			\hspace{-0.5 in}	$^{4}$Department of Geography, University of North Carolina, Chapel Hill, North Carolina, U.S.A.\\	\vspace{-0.15 in}
} 

\begin{document}

\begin{abstract}	
Inferring causal effects from an observational study is challenging because participants are not randomized to treatment. Observational studies in infectious disease research present the additional challenge that one participant's treatment may affect another participant's outcome, i.e., there may be interference. In this paper recent approaches to defining causal effects in the presence of interference are considered, and new causal estimands designed specifically for use with observational studies are proposed. Previously defined estimands target counterfactual scenarios in which individuals independently select treatment with equal probability. However, in settings where there is interference between individuals within clusters, it may be unlikely that treatment selection is independent between individuals in the same cluster. The proposed causal estimands instead describe counterfactual scenarios in which the treatment selection correlation structure is the same as in the observed data distribution, allowing for within-cluster dependence in the individual treatment selections. These estimands may be more relevant for policy-makers or public health officials who desire to quantify the effect of increasing the proportion of treated individuals in a population. Inverse probability-weighted estimators for these estimands are proposed. The large-sample properties of the estimators are derived, and a simulation study demonstrating the finite-sample performance of the estimators is presented. The proposed methods are illustrated by analyzing data from a study of cholera vaccination in over 100,000 individuals in Bangladesh. 
\end{abstract}
 


\maketitle


\section{Introduction} \label{sec:Introduction}

Inferring causal effects from an observational (i.e., non-randomized or non-experimental) study is challenging because participants may select their own treatment. Observational studies in many settings such as infectious disease research present the additional challenge that one individual's treatment may have an effect on another individual's outcome, i.e., there may be interference \citep{cox1958planning}. For example, whether one individual is administered a vaccine may affect whether another individual develops disease from some infectious pathogen. In certain settings it may be reasonable to assume that individuals can be partitioned into clusters such that there may be interference among individuals within a single cluster, yet no interference between individuals in distinct clusters. \citet{Sobel2006RandomizedStudies} described this assumption as ``partial interference''; here this assumption is referred to as ``clustered interference.'' Clusters might entail households, classrooms, geographical areas, or other hierarchical structures. Several types of treatment effects (i.e., causal estimands) have been proposed for the setting where there may be clustered interference; e.g., see \citet{halloran1995causal}, \citet{Hudgens2008Toward} and \citet{Tchetgen2012OnCausalInference}.

Methods have been developed for inference about these causal effects from observational studies \citep{Tchetgen2012OnCausalInference,PerezHeydrich2014Assessing,liu2016inverse}. One drawback of the treatment effects targeted by these methods is that these causal estimands describe counterfactual scenarios in which individuals select treatment independently and with the same probability. However, in settings where interference within clusters is plausible, it may be unlikely that treatment selection among individuals in the same cluster is independent \citep{liu2016inverse}. For instance, suppose a public health policy-maker is interested in the effect of seasonal influenza vaccination on risk of influenza-like illness in households. In this case, one might expect positive correlation between the vaccination statuses of individuals in the same household. Thus, drawing inference to a counterfactual scenario in which individuals are administered vaccines independently may not be of public health relevance. In this paper new causal estimands are proposed for observational studies where there may be clustered interference; these estimands describe counterfactual scenarios in which the treatment selection correlation structure is the same as that in the observed data distribution. By considering scenarios that exhibit within-cluster dependence in the individual treatment selections, the proposed estimands may be more relevant for policy-makers or public health officials who are interested in quantifying the effect of increasing the proportion of treated individuals in a population.

The outline of the remainder of this paper is as follows. In Section~\ref{sec:PotentialOutcomes} the potential outcomes framework and interference are discussed. The proposed causal estimands are introduced in Section~\ref{sec:Estimands}. Identification assumptions for the target estimands are presented in Section~\ref{sec:IDable}. In Section~\ref{sec:Estimators} inverse probability-weighted estimators are introduced; the estimators are shown in 
\hyperlink{sec:EEqn_var}{Appendix A} to be consistent and asymptotically Normal. Simulations in Section~\ref{sec:Simulations} show that the proposed estimators are empirically unbiased and that their confidence intervals attain nominal coverage levels in finite samples. The proposed methods are illustrated in Section~\ref{sec:Cholera} by analyzing data from a study of cholera vaccination in over 100,000 individuals in Matlab, Bangladesh. Section~\ref{sec:Discussion} concludes with a discussion.

\section{Counterfactuals and Interference} \label{sec:PotentialOutcomes}

Consider a super-population of clusters of individuals. For each cluster let $N$ be the number of individuals in the cluster, $A=(A_1,A_2,\ldots,A_N)$ where $A_j$ denotes the binary treatment indicator for individual $j$ in the cluster, and $Y=(Y_1,Y_2,\ldots,Y_N)$ where $Y_j$ is the outcome of interest for individual $j$. For example, $Y_j$ might indicate whether or not individual $j$ experienced the outcome after some suitable follow-up period after treatment exposure status was observed.

Assuming clustered interference, the potential outcome for an individual may depend on the individual's own treatment exposure status as well as on the treatment statuses of others in the same cluster. However, any individual's potential outcomes are assumed to be unaffected by the treatment exposures of individuals in different clusters. Let $\mathcal{A}(N)$ be the set of all vectors with $N$ binary entries such that $a=(a_1, a_2, \dots, a_N) \in \mathcal{A}(N)$ is a vector of potential treatment statuses for a cluster of size $N$. Let $Y_j(a)$ be the potential outcome for unit $j$ in the cluster if, possibly counter to fact, the cluster had been exposed to $a\in \mathcal{A}(N)$. In the absence of interference, $Y_j(a) = Y_j(a')$ whenever $a_j={a}'_j$ for $a, a' \in \mathcal{A}(N)$. However, assuming no interference when interference is present may result in biased estimates of causal effects. Throughout this paper clustered interference is assumed.

\section{Estimands} \label{sec:Estimands} 

Our goal is to draw inference about the difference in expected outcomes arising from population-level policies which change the distribution of treatment. Typical treatment effect estimands compare the policy where all individuals receive treatment (i.e.,  $A=(1,1,\dots, 1)$ with probability 1) with the policy where all individuals are not treated (i.e., $A=(0,0,\ldots,0)$ with probability 1). Here we consider more general policies where individuals receive treatment according to some  probability. \citet{munoz2012population} refer to such policies as ``stochastic interventions.'' For example, we might consider the policy where individuals select treatment with probability $1/2$. In general, let $\alpha$ denote the policy under which the probability an individual is treated equals $\alpha$, for $\alpha \in [0,1]$. That is,
\begin{equation} \label{def:alpha} 
{\Pr}_{\alpha }(A_j={1})=\alpha,
\end{equation}
where the subscript in ${\Pr}_{\alpha}(\cdot)$ indicates that the probability is with respect to the counterfactual scenario in which the policy $\alpha$ is implemented.

For $a \in \mathcal{A}(N)$, define $\omega(a,N,\alpha )= {\Pr}_{\alpha }(A=a|N)$ to be the marginal probability under policy $\alpha$ that a cluster of $N$ individuals experiences treatment status $a$. Let $\overline{Y}(a) = N^{-1}\sum_{j=1}^{N} Y_j(a) $ denote the average potential outcome in a cluster if the cluster had been exposed to $a$. The expected potential outcome under $\alpha$ for a single cluster of $N$ individuals is defined to be $\overline{Y}(\alpha ) = \sum_{a\in \mathcal{A}(N)} \overline{Y}(a) \omega(a, N,\alpha )$. In other words, $\overline{Y}(\alpha )$ is the expected average potential outcome for the cluster in the counterfactual scenario in which $\alpha $ is implemented.

Define the population mean outcome under $\alpha$ to be $\mu(\alpha ) = \Eval \{ \overline{Y}(\alpha ) \}$, where the expected value is taken over all clusters in the super-population. The overall effect is defined to be $\OVE(\alpha , \alpha' ) = \mu(\alpha ) - \mu(\alpha' )$, which represents the difference in expected potential outcomes under policy $\alpha$ versus policy $\alpha' $. The overall effect is defined here as a difference in mean potential outcomes, but could instead be defined as a ratio or some other contrast \citep{liu2016inverse}. Below in Section~\ref{sec:Estimators}, methods are considered for drawing inference about the target estimands, $\mu(\alpha)$ and $\OVE(\alpha, \alpha')$, for different policies $\alpha$ and $\alpha'$.

\subsection{Spillover metrics}\label{sec:spillover}

In addition to the target estimands, it may also be of interest to consider potential outcomes among only the untreated individuals within a cluster. Let $\overline{Y}_t(a)= \{\sum_{j=1}^{N} I(a_{j}=t)\}^{-1} \sum_{j=1}^{N} Y_j(a)I(a_j=t)$ for $t=0,1$. In words, $\overline{Y}_0(a)$ is the average potential outcome among the untreated individuals within the cluster; likewise $\overline{Y}_1(a)$ is the average potential outcome among the treated individuals within the cluster. In the special case where $a=(1-t, 1-t, \dots, 1-t)$, define $\overline{Y}_t(a)=0$ for each of $t=0,1$. Denote the population mean potential outcomes when untreated to be $\mu_0(\alpha ) = \Eval\{ \sum_{a\in \mathcal{A}(N)} \overline{Y}_0(a) \omega(a, N,\alpha )\}$. The spillover effect when untreated is defined to be the difference in population mean potential outcomes when untreated under policy $\alpha$ versus $\alpha'$, i.e., $\SE_0(\alpha , \alpha' ) = \mu_0(\alpha ) - \mu_0(\alpha' )$. Similarly, let $\mu_1(\alpha)=\Eval\{ \sum_{a\in \mathcal{A}(N)} \overline{Y}_1(a) \omega(a, N,\alpha )\}$, and define $\SE_1(\alpha , \alpha' ) = \mu_1(\alpha ) - \mu_1(\alpha' )$ to be the spillover effect when treated.

\subsection{Relation to existing estimands}\label{sec:TV}

Consider a policy in which all individuals in a cluster are exposed to treatment independently with the same probability; \citet{Tchetgen2012OnCausalInference} refer to this as a ``type B parameterisation.'' For $\alpha \in [0,1]$, let $\omegaTV(a,N,\alpha)= \prod_{j=1}^{N} \alpha^{a_j}(1-\alpha)^{1-a_j}$ denote the counterfactual probabilities under such a type B policy. Likewise, let $\muTV(\alpha) = \Eval\{\sum_{a \in \mathcal{A}(N) }\overline{Y}(a)\omegaTV(a,N,\alpha)\}$ be the population mean potential outcome for the type B policy with parameter $\alpha$, and define the overall effect with respect to two type B policies to be $\OVETV(\alpha, \alpha') = \muTV(\alpha) - \muTV(\alpha')$.

A type B policy is a special case of the policies of interest that corresponds to the counterfactual scenarios in which treatment exposure is uncorrelated. The estimands proposed in this paper can thus be seen as a generalization of the type B estimands, as the type B policies describe only the limiting counterfactual scenarios in which there is no within-cluster dependence of individual treatment selections. In general, $\omega(a,n,\alpha) \neq \omegaTV(a,n,\alpha)$ and the corresponding policies, estimands, and interpretations differ. In the data analysis of the cholera vaccine study in Section~\ref{sec:Cholera}, estimates of the type B estimands will be presented for comparison to the estimates of the proposed estimands.

\section{Identifiability}\label{sec:IDable}

The counterfactual probabilities $\omega(a,n,\alpha )$ are not identifiable without additional assumptions. Below we assume no unmeasured confounders and parametric models of the conditional distribution of treatment given covariates.

Let there be a random sample of $i=1,\dots, M$ clusters, 
and denote by $O_i=\{N_i, L_i, A_i, Y_i \}$ the observed values of the random variables for cluster $i$, where $L_i$ is a vector of baseline (i.e., pre-treatment) variables. The subscript $i$ is dropped for notational simplicity when not needed. Assume exchangeability conditional on the baseline variables at the cluster level:
\begin{equation*} \label{def:NUC}
Y(a) \;\perp \;A \; | \; L, N \text{ for any }a \; \in\; \mathcal{A}(N).
\end{equation*}
In addition assume cluster-level positivity:
\begin{equation*} \label{def:OverlapAssumption}
\Pr(A=a|L, N )>0 \text{  for any } a \in \mathcal{A}(N).
\end{equation*}
Following \citet{Tchetgen2012OnCausalInference}, \citet{PerezHeydrich2014Assessing}, and \citet{liu2016inverse}, assume the following mixed effects logistic regression model for treatment:
\begin{equation} \label{eq:LMEM}
\Pr(A=a|L,N) = \int \prod_{j=1}^N  \expit(\beta_0+\beta_1L_{j} + b )^{a_j} \bigl\{1-\expit(\beta_0+\beta_1L_{j} + b )\bigr\}^{(1-a_j)} d\Phi(b ;\sigma),
\end{equation}
where $\expit(x)= \{1+\exp(-x)\}^{-1}$ is the inverse-logit function, and $b $ denotes a random intercept for cluster which is assumed to follow a Normal distribution with mean zero, standard deviation $\sigma$, and distribution function $\Phi(\cdot)$. We refer to $\Pr(A=a|L,N)$ as a cluster propensity score \citep{rosenbaum1983central}. These conditional probabilities describe the relationship between observed treatment and covariates; unlike in the case where no interference is assumed, each of these is a scalar probability of the joint exposure statuses of all individuals within the cluster. 

In addition, assume under counterfactual policy $\alpha$ that 
\begin{equation*} \label{eq:CFLMEM}
{\Pr}_{\alpha }(A=a|L,N) = \int \prod_{j=1}^N \expit(\gamma_{0\alpha } + \gamma_{1\alpha }L_j +  b )^{a_j} \bigl\{1-\expit(\gamma_{0\alpha } + \gamma_{1\alpha }L_j +  b )\bigr\}^{(1-a_j)} d\Phi(b ;\phi_{\alpha }),
\end{equation*}
where the random intercept follows a Normal distribution with mean zero and standard deviation $\phi_{\alpha }$. The model parameters in the counterfactual scenario in general may differ from the parameters in the factual scenario. We similarly refer to ${\Pr}_{\alpha}(A=a|L,N)$ as a counterfactual cluster propensity score, as these conditional probabilities describe the relationship between treatment and covariates in the counterfactual scenario in which $\alpha$ is implemented.
 
The parameters $(\beta_0, \beta_1, \sigma)$ in \eqref{eq:LMEM} are identifiable from the observable data. However, the parameters $(\gamma_{0\alpha }, \gamma_{1\alpha } , \phi_{\alpha })$, counterfactual cluster propensity scores ${\Pr}_{\alpha}(A=a| L, N)$, and counterfactual probabilities $\omega(a,n, \alpha)$ are not identifiable without additional assumptions. It is assumed here that $\Pr(L) = \Pr_{\alpha }(L)$, i.e., the different policies do not affect the covariate distribution. Also assume that $\sigma= \phi_{\alpha }$, i.e., the parameter governing correlation is not affected by different policies. Additionally assume $\beta_1 = \gamma_{1\alpha }$. In brief, this supposes that the ranking of individuals within each cluster by the conditional probability of treatment is preserved across factual and counterfactual scenarios; further discussion regarding this assumption is provided in Section~\ref{sec:Discussion}. Under the above assumptions, \eqref{def:alpha} implies 
\begin{equation}\label{eq:alpha_ID}
\alpha = \int  \left\{N^{-1}\sum\limits_{j=1}^{N} \int  \expit(\gamma_{0\alpha}  + \beta_1L_{j} + b)d\Phi(b;\sigma)\right\} dF_L, 
\end{equation}
so the counterfactual model intercept parameter $\gamma_{0\alpha} $ and thus the counterfactual cluster propensity scores are identifiable. It follows that the counterfactual probabilities $\omega(a,n,\alpha ) = \Eval_{L}\{{\Pr}_{\alpha }(A = a | L , N=n)\}$ are also identifiable from the observable data.

\section{Inference}\label{sec:Estimators} 

Following \citet{Tchetgen2012OnCausalInference} and \citet{PerezHeydrich2014Assessing}, consider the following inverse probability-weighted (IPW) estimator of $\mu(\alpha)$:
\begin{equation} \label{est:mu_hat}
\widehat{\mu}(\alpha )= {M}^{-1}\sum\limits_{i=1}^{M}   \frac{\overline{Y}_i   {\omega}(A_i,N_i,\alpha )}{{\Pr}(A_i|L_i, N_i)}, 
\end{equation}
where $\overline{Y}_i = N_i^{-1}\sum_{j=1}^{N_i} Y_{ij}$. The inverse probability-weight for cluster $i$ is the reciprocal of the cluster propensity score; these and the counterfactual probabilities are unknown in an observational study and must be estimated from data. 

Under the assumptions in Section~\ref{sec:IDable}, a logistic mixed effects model is fit to the data, and the model parameters $(\beta_0, \beta_1, \sigma)$ can be estimated by maximum likelihood. Then, the fitted parameters $(\hat{\beta}_0, \hat{\beta}_1, \hat{\sigma})$ are substituted into \eqref{eq:LMEM} to obtain an estimate of each cluster's propensity score. For each policy $\alpha$ , $\hat{\gamma}_{0\alpha}$ solves equation \eqref{eq:alpha_ID}, with $F_L$ replaced by its empirical distribution. That is, $\alpha\ =  M^{-1}\sum_{i=1}^M N_i^{-1} \sum_{j=1}^{N_i} \int \expit(\gamma_{0\alpha}  + \hat{\beta}_{1}L_{ij} + b_i) d\Phi(b_i; \hat{\sigma})$ is solved to obtain $\hat{\gamma}_{0\alpha}$. The counterfactual cluster propensity scores for cluster $i$ and treatments $a\in \mathcal{A}(N_i)$ are estimated by substitution, e.g.,
\begin{equation*} \label{eq:CCFP} 
\widehat{\Pr}_{\alpha }(A_i = a  | L_i, N_i) = \int \prod_{j=1}^{N_i}
\expit(\hat{\gamma}_{0\alpha} + \hat{\beta}_1L_{ij} +  b_i )^{a_j}\bigl\{1-\expit(\hat{\gamma}_{0\alpha} + \hat{\beta}_1L_{ij} +  b_i )\bigr\}^{(1-a_j)}d\Phi(b_i; \hat{\sigma}).
\end{equation*}

Assume the ordering of individuals in clusters to be uninformative, and so $\omega(a,n,\alpha )=\omega(a',n,\alpha )$ whenever $f(a)=f(a')$ for any two $a, a' \in \mathcal{A}(n)$ where $f(a)=\sum_{j=1}^n a_j$. Thus the maximum number of unique counterfactual probabilities is reduced; see Appendices \hyperlink{sec:EEqn_var}{A} and \hyperlink{sec:AppB}{B} for further details. Let $\mathcal{A}(n,s) = \{a \in \mathcal{A}(n) \, | \, f(a) = s\} $ such that $|\mathcal{A}(n,s)|=\binom{n}{ s}$, and define $\omega(s,n,\alpha ) = \sum_{a\in \mathcal{A}(n,s)} \omega(a,n,\alpha ) $ for $s=0,1,\dots, n$. Estimate the counterfactual probabilities for any cluster $i$ by $\widehat{\omega}(A_i,N_i, \alpha ) = \binom{N_i}{ f(A_i)}^{-1} \widehat{\omega}(f(A_i),N_i, \alpha )$, where for any triplet $(s,n,\alpha)$,
\begin{equation*} 
\widehat{\omega}(s,n, \alpha ) =   \left\{  
\sum\limits_{i=1}^M I(N_i=n) \right\}^{-1} \sum\limits_{a  \in \mathcal{A}(n,s)} \sum\limits_{i=1}^M \widehat{\Pr}_{\alpha }(A_i=a |L_i , N_i)I(N_i=n).
\end{equation*} 
These estimates, along with the estimated cluster propensity scores, are substituted into \eqref{est:mu_hat} to calculate $\widehat{\mu}(\alpha )$. The estimator $\widehat{\OVE}(\alpha , \alpha' ) = \widehat{\mu}(\alpha) - \widehat{\mu}(\alpha')$ can be obtained in a similar manner. For $t=0,1$, the estimators $\widehat{\mu}_t(\alpha )$ and $\widehat{\SE}_t(\alpha , \alpha')$ are defined similarly using the outcomes $\overline{Y}_{t,i}=\{\sum_{j=1}^{N_i} I(A_{ij}=t)\}^{-1} \sum_{j=1}^{N_i} Y_{ij}I(A_{ij}=t)$, where $\overline{Y}_{t,i}=0$ in the case when $A_{ij}=1-t$ for all $j=1, \dots, N_i$.

In 
\hyperlink{sec:EEqn_var}{Appendix A} these estimators are shown to be consistent and asymptotically Normal using standard large-sample estimating equation theory \citep{Stefanski2002Calculus}. Wald-type confidence intervals (CIs) can be constructed using the empirical sandwich estimators of the asymptotic variances.

The estimators described above may be computationally challenging in practice as the estimator $\widehat{\omega}(a,n,\alpha)$ calls a numerical integration technique for each of the $\binom{n}{f(a)}$-many vectors in $\mathcal{A}(n,f(a))$. An approximate technique is proposed that uses only a randomly sampled subset of the vectors to decrease computation time. For each $s=0,1,\dots, n$, define $\mathcal{A}(n,s,k)$ to be a subset of exactly $k_{s,n} = \min\{k, \binom{n}{s}\}$ vectors constructed from a simple random sample from $\mathcal{A}(n, s)$,  
where $k>1$ is chosen by the investigator. Now estimate the counterfactual probabilities by $\widehat{\omega}(a,n,\alpha ,k ) =  \binom{n}{ f(a)}^{-1} \widehat{\omega}(f(a),n, \alpha ,k)$, where
\begin{equation*}
\widehat{\omega}(s,n, \alpha ,k) =
\left\{ \sum\limits_{i=1}^M I(N_i=n) \right\}^{-1} k_{s,n}^{-1}\binom{n}{ s} 
 \sum\limits_{a  \in \mathcal{A}(n,s,k)} \sum\limits_{i=1}^M \widehat{\Pr}_{\alpha }(A_i=a |L_i, N_i)I(N_i=n)
\end{equation*}  
evaluates over only the $k_{s,n}$-many sub-sampled vectors and up-weights by $k_{s,n}^{-1}\binom{n}{ s}$.
 
Replacing $\widehat{\omega}(a,n, \alpha )$ in $\widehat{\mu}(\alpha )$ with $\widehat{\omega}(a,n, \alpha ,k)$ results in an estimator which we denote $\widehat{\mu}(\alpha ,k)$. Making analogous replacements, define $\widehat{\OVE}(\alpha, \alpha', k)$, as well as $\widehat{\mu}_t(\alpha ,k)$ and $\widehat{\SE}_t(\alpha , \alpha' ,k)$ for $t=0,1$. These estimators are evaluated in a simulation study in Section~\ref{sec:Simulations} and are employed in the data analysis of the cholera vaccine study in Section~\ref{sec:Cholera}. In practice, specification of the value of $k$ may be a compromise between less approximation (larger $k$) and faster computation (smaller $k$). An extension of this method is to specify different values of $k$ to estimate distinct counterfactual probabilities, which is outlined in \hyperlink{sec:EEqn_var}{Appendix A}.


\section{Simulations} \label{sec:Simulations}
 
A simulation study was carried out on 1000 datasets to demonstrate the finite-sample performance of the proposed estimators. To generate each dataset, the following steps were carried out for each of $M=125$ clusters: 

\begin{enumerate}[i.]
	\item \label{step1} The number of individuals in the cluster $N_i$ was simulated such that $\Pr(N_i=8)= 0.4$, $\Pr(N_i=22)=0.35$, and $\Pr(N_i=40)=0.25$.
	
	\item \label{step2} Covariates for each individual $j=1,\dots, N_i$ in cluster $i$ were simulated to be $L_{1ij} \sim N(40,5)$ and $L_{2ij} \sim N(L_{*i},0.2)$, where $L_{*i} \sim N(6,1)$ was a cluster-level random variable.
	
	\item \label{step3} Treatment status $A_{ij}$ was simulated from a Bernoulli distribution with mean $\Pr(A_{ij}=1|L_{ij},b_i) =\expit(\beta_0 + \beta_1L_{1ij} +\beta_2L_{2ij}+b_i)$ where $b_i\sim N(0, \sigma)$ was a cluster-level random intercept and $(\beta_0, \beta_1, \beta_2,\sigma )= (0.75,$ $-0.015, -0.025, 0.75)$.
	
	\item \label{step4} The outcome $Y_{ij}$ for each individual $j$ was simulated from a Bernoulli distribution with mean  
	$\Pr(Y_{ij}=1|A_{i}, L_{ij}) = \expit(0.1 - 0.05L_{1ij} + 0.5L_{2ij} -0.5A_{ij} + 0.2g(A_{i,-j}) -0.25A_{ij}g(A_{i,-j})  )$,
	where $g(A_{i,-j}) = (N_i-1)^{-1}\sum_{j'\neq j} A_{ij'}$.
\end{enumerate}

A logistic mixed effects model was fit with a random intercept and main effects for $L_1$ and $L_2$, i.e., the propensity score models were correctly specified. 
To determine the performance of the estimators that use the greatest degree of sub-sampling approximation, $k=1$ was chosen. The asymptotic variance of the estimators was estimated with the empirical sandwich variance estimator as described in 
\hyperlink{sec:EEqn_var}{Appendix A}, from which Wald-type 95\% CIs were constructed. 

True values of the estimands for policies $\alpha \in \{0.4, 0.5, 0.55\}$ were determined empirically, using the same data generating process outlined above in steps~\ref{step1}-\ref{step2} and analogues to steps~\ref{step3}-\ref{step4}.
The process is described here briefly, with more details provided in \hyperlink{sec:AppC}{Appendix C}. For each $\alpha$, the counterfactual probabilities $\omega(a,n,\alpha)$ were determined by replacing $\beta_0$ with $\gamma_{0\alpha}$ in step~\ref{step3} to generate treatment vectors under policy $\alpha$ for $10^8$ clusters, where $\gamma_{0\alpha}$ was determined by solving \eqref{eq:alpha_ID} with $F_L$ approximated by its empirical distribution over $10^7$ clusters. Potential outcomes were generated for $10^8$ clusters via the causal model analogous to the regression model specified in step~\ref{step4}, and these were combined with the counterfactual probabilities to determine the true values of $\mu(\alpha)$, $\OVE(\alpha, \alpha')$, and $\mu_t(\alpha)$ and $\SE_t(\alpha, \alpha')$ for $t=0,1$. 

An empirical comparison of true values of $\omega(a,n,\alpha )$ arising from this simulation study and the true values of $\omegaTV(a,n,\alpha )$ for the type B policies is provided in Figure~\ref{fig:omega_pi_compare2017-08-07} in \hyperlink{sec:AppC}{Appendix C}.

The IPW estimates from each dataset were compared to the true estimand values determined above; a summary of these results is presented below in Table~\ref{table:Sims_small}. The average bias of the estimators was negligible. The average of the estimated asymptotic standard errors was approximately equal to the empirical Monte Carlo standard error. The Wald-type 95\% CIs contained the true parameter values for approximately 95\% of the simulated datasets. Thus, the estimators performed well in this simulation study.

\begin{table}[t]   
	\centering
	\caption{
		Summary of results from simulation study described in Section~\ref{sec:Simulations}. 
		Truth denotes the true value of the estimand targeted by the estimator; Bias denotes the average bias of the IPW estimates over the 1000 datasets; Cov\% denotes the empirical coverage of Wald-type 95\% CIs; ASE denotes the average of the estimated sandwich standard errors times $100$; ESE denotes the empirical standard error times $100$; SER denotes the ratio of ASE divided by ESE; $\alpha_1 = 0.4$, $\alpha_2 = 0.5$, and $\alpha_3 = 0.55$.
		} 
	\label{table:Sims_small}
		\begin{tabular}{lrrrrrrr}
			\hline
			Estimator & Truth  &  Bias & Cov\% &  ASE   &  ESE  & SER  \\ 
			\hline
			$\widehat{\mu}(\alpha_1,k=1)$ 		&  0.662 & -0.003 & 94.3\% & 1.88 & 1.84 & 1.02 \\ 
			$\widehat{\mu}(\alpha_2,k=1)$ 		&  0.651 &  0.000 & 95.5\% & 1.63 & 1.53 & 1.06 \\ 
			$\widehat{\mu}(\alpha_3,k=1)$ 		&  0.645 &  0.001 & 96.4\% & 1.65 & 1.55 & 1.07 \\ 
			$\widehat{\OVE}(\alpha_2, \alpha_1,k=1)$  & -0.011 &  0.003 & 97.2\% & 1.08 & 0.96 & 1.13 \\ 
			$\widehat{\OVE}(\alpha_3, \alpha_1,k=1)$  & -0.017 &  0.004 & 97.4\% & 1.44 & 1.34 & 1.08 \\ 
			$\widehat{\OVE}(\alpha_3, \alpha_2,k=1)$  & -0.006 &  0.001 & 97.4\% & 0.53 & 0.44 & 1.21 \\ \hline 
			$\widehat{\mu}_0(\alpha_1 ,k=1)$ 	&  0.712 & -0.002 & 95.2\% & 2.10 & 2.02 & 1.04 \\ 
			$\widehat{\mu}_0(\alpha_2 ,k=1)$ 	&  0.711 & -0.001 & 95.7\% & 2.15 & 2.02 & 1.07 \\ 
			$\widehat{\mu}_0(\alpha_3 ,k=1)$ 	&  0.709 & -0.001 & 95.3\% & 2.46 & 2.35 & 1.05 \\ 
			$\widehat{\SE}_0(\alpha_2, \alpha_1 ,k=1)$ & -0.001 &  0.001 & 95.8\% & 1.33 & 1.20 & 1.11 \\ 
			$\widehat{\SE}_0(\alpha_3, \alpha_1 ,k=1)$ & -0.003 &  0.001 & 94.7\% & 1.93 & 1.86 & 1.04 \\ 
			$\widehat{\SE}_0(\alpha_3, \alpha_2 ,k=1)$ & -0.002 & 0.000 & 94.8\% & 0.79 & 0.72 & 1.10 \\ \hline
			$\widehat{\mu}_1(\alpha_1 ,k=1)$ 	&  0.573 & 0.007 & 94.2\% & 3.04 & 3.09 & 0.99 \\ 
			$\widehat{\mu}_1(\alpha_2 ,k=1)$ 	&  0.581 & 0.004 & 95.0\% & 2.25 & 2.24 & 1.01 \\ 
			$\widehat{\mu}_1(\alpha_3 ,k=1)$ 	&  0.582 & 0.001 & 95.3\% & 2.10 & 2.07 & 1.01 \\ 
			$\widehat{\SE}_1(\alpha_2, \alpha_1 ,k=1)$ &  0.008 &  0.003 & 94.9\% & 1.51 & 1.46 & 1.04 \\ 
			$\widehat{\SE}_1(\alpha_3, \alpha_1 ,k=1)$ &  0.009 &  0.005 & 95.2\% & 2.02 & 1.98 & 1.02 \\ 
			$\widehat{\SE}_1(\alpha_3, \alpha_2 ,k=1)$ &  0.002 &  0.002 & 96.4\% & 0.65 & 0.57 & 1.13 \\ 
			\hline
			\hline 
			 
		\end{tabular}  
\end{table}


\section{Analysis of Cholera Vaccine Trial in Matlab, Bangladesh} \label{sec:Cholera}

The proposed methods are illustrated in the following analysis of a cholera vaccine study in Matlab, Bangladesh, which featured both an experimental and a non-experimental component \citep{ali2005herd, ali2009modeling}. Included in the study were 121,975 women (aged 15 years and older) and children (aged 2-15 years) from 6,415 baris (i.e., households of patrilineally-related individuals). These individuals were eligible to participate in the experimental component of the study, in which each individual was randomized with equal probability to one of three treatment arms: B subunit-killed whole cell oral cholera vaccine, killed whole cell-only oral cholera vaccine, or placebo. Individuals who did not participate did not receive either version of active treatment. The study collected endpoint data of cholera infection on all individuals, even those who did not participate in the experimental component. Since participation was not controlled by study design and nearly two-fifths of all individuals declined to participate, there was a notable non-experimental component to the study, and potential for confounding exists when analyzing the endpoint data. 

\begin{figure}
\centering
\includegraphics[width=1\linewidth]{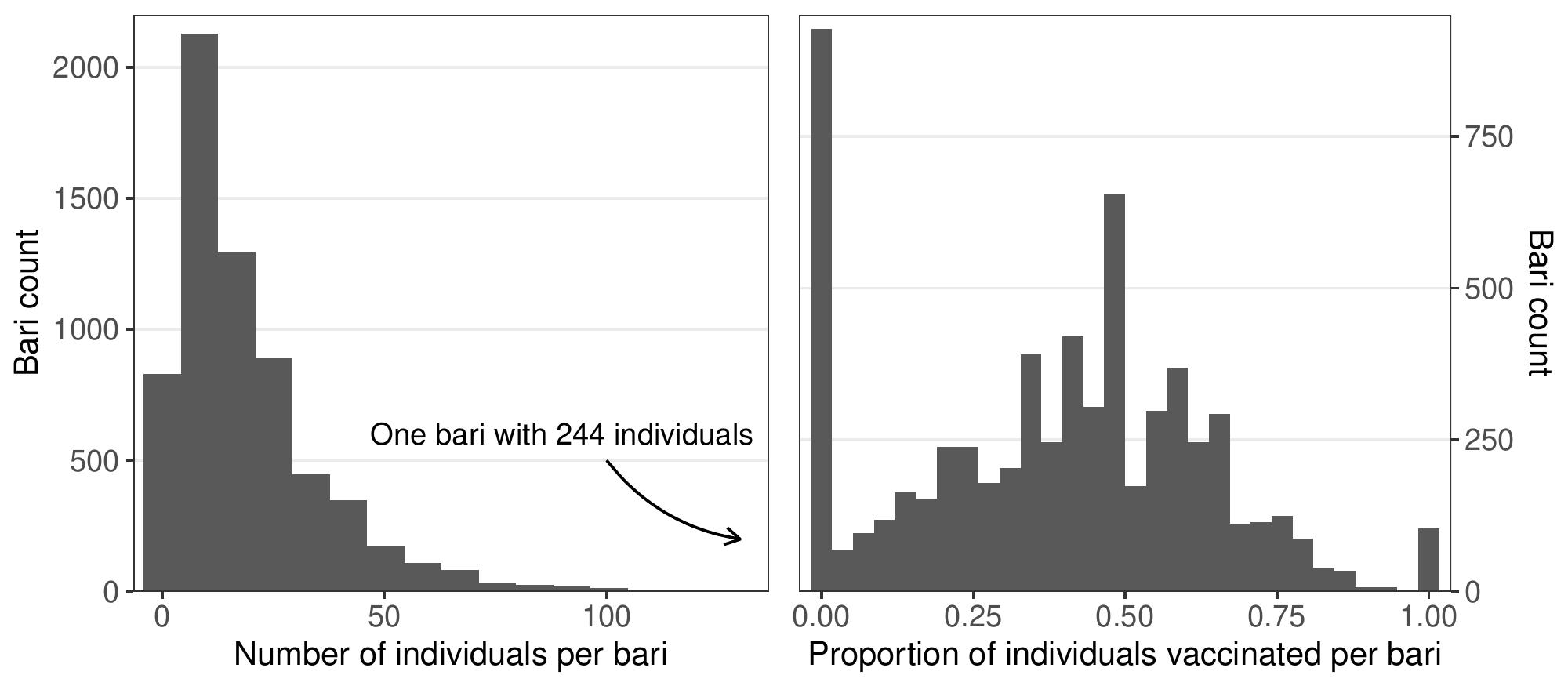}
\caption{Empirical distribution of individuals per cluster (bari), and proportion of individuals vaccinated per cluster.}
\label{fig:plots2and3}
\end{figure}

As in \citet{PerezHeydrich2014Assessing}, any individual who received at least two doses of either of the two cholera vaccines was considered to be treated, and otherwise was considered to be untreated. Clustered interference was assumed at the level of the bari as there is evidence that transmission of cholera often takes place within baris \citep{ali2005herd}. Figure~\ref{fig:plots2and3} illustrates the empirical distributions of the number of individuals and of the treatment coverage within the baris. 

Cluster-level conditional exchangeability and positivity were assumed to hold conditional on age and distance from the bari to the nearest river. A logistic mixed effects model was fit, regressing the indicator that an individual obtained treatment on the individual's age and river distance with a random intercept for the bari in which the individual lived. Included in the mixed effects logistic regression model was a linear term for distance (in kilometers) and linear and quadratic terms for age (centered, in decades). All the assumptions for identifiability as discussed in Section~\ref{sec:IDable} were assumed. The IPW estimators were computed with $k=3$, and Wald-type CIs were constructed from the empirical sandwich variance estimator.

\begin{figure}
	\centering
	\includegraphics[width=0.6\linewidth]{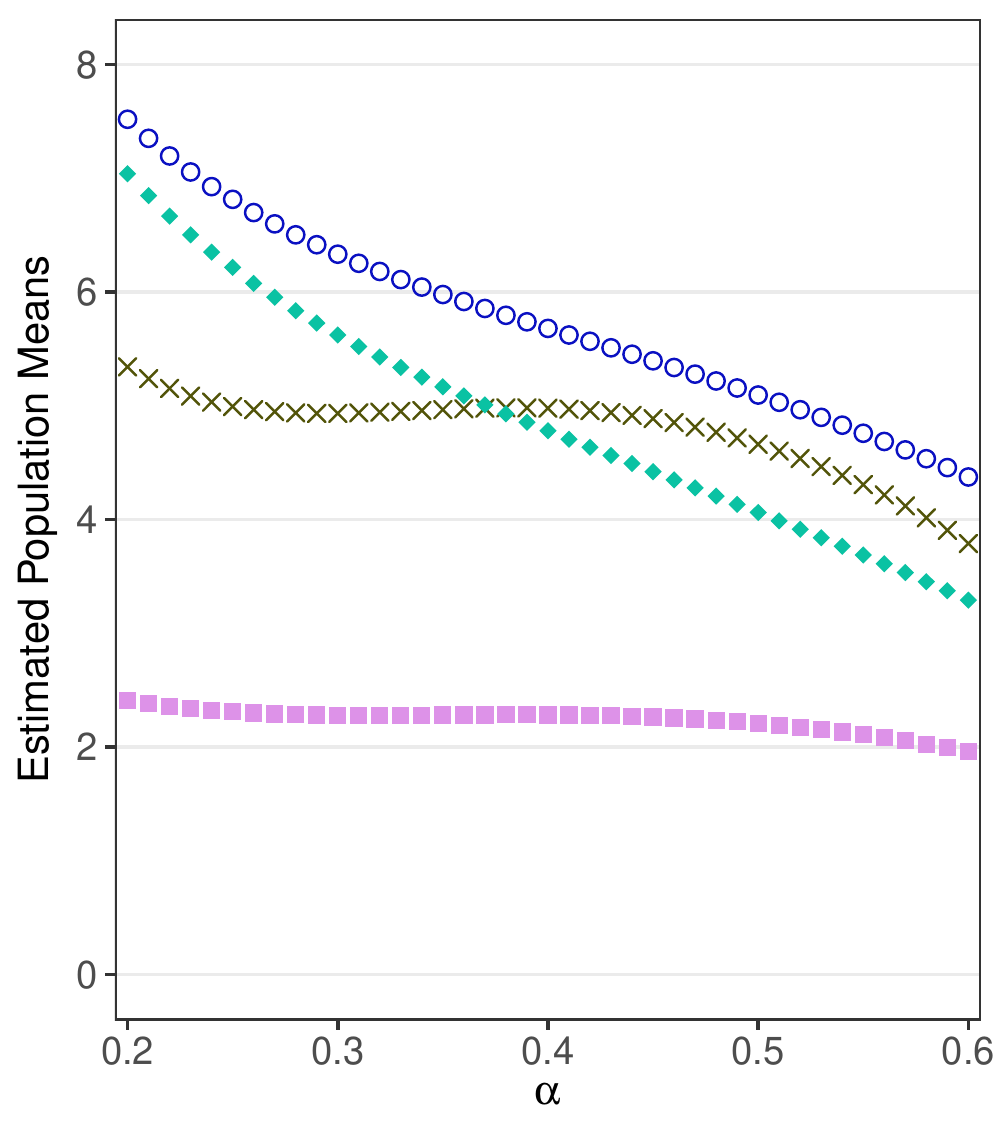}
	\caption{
		Estimates of the population mean estimands from the analysis of the Matlab cholera vaccine study. 
		The light green diamonds indicate $\widehat{\mu}(\alpha,k=3)$. 
		The dark blue circles indicate $\widehat{\mu}_0(\alpha, k=3)$, and the light pink squares indicate $\widehat{\mu}_1(\alpha, k=3)$.
		The dark brown $\times$'s indicate $\hmuTV(\alpha)$, which target the type B estimands from \citet{Tchetgen2012OnCausalInference}. 
		All estimates are multiplied by 1000. 
		This figure appears in color in the electronic version of this article.
		}
	\label{fig:MCVT_mus_plot}
\end{figure}

Figure~\ref{fig:MCVT_mus_plot} depicts point estimates of the population mean estimands over policies ranging from $\alpha=0.2$ to $\alpha=0.6$. Estimates are presented in units of one case of cholera infection per 1000 individuals per year. Estimates of $\mu_1(\alpha)$ were relatively invariant to $\alpha$, suggesting minimal spillover effects when an individual is vaccinated. In contrast, estimates of $\mu_0(\alpha)$ decreased noticeably as $\alpha$ increased, suggesting a protective spillover effect when an individual is not vaccinated. The estimates of $\mu(\alpha)$ similarly suggest lower risk of cholera infection at the population level for policies with greater levels of vaccine coverage.

\begin{figure}
	\centering
	\includegraphics[width=1\linewidth]{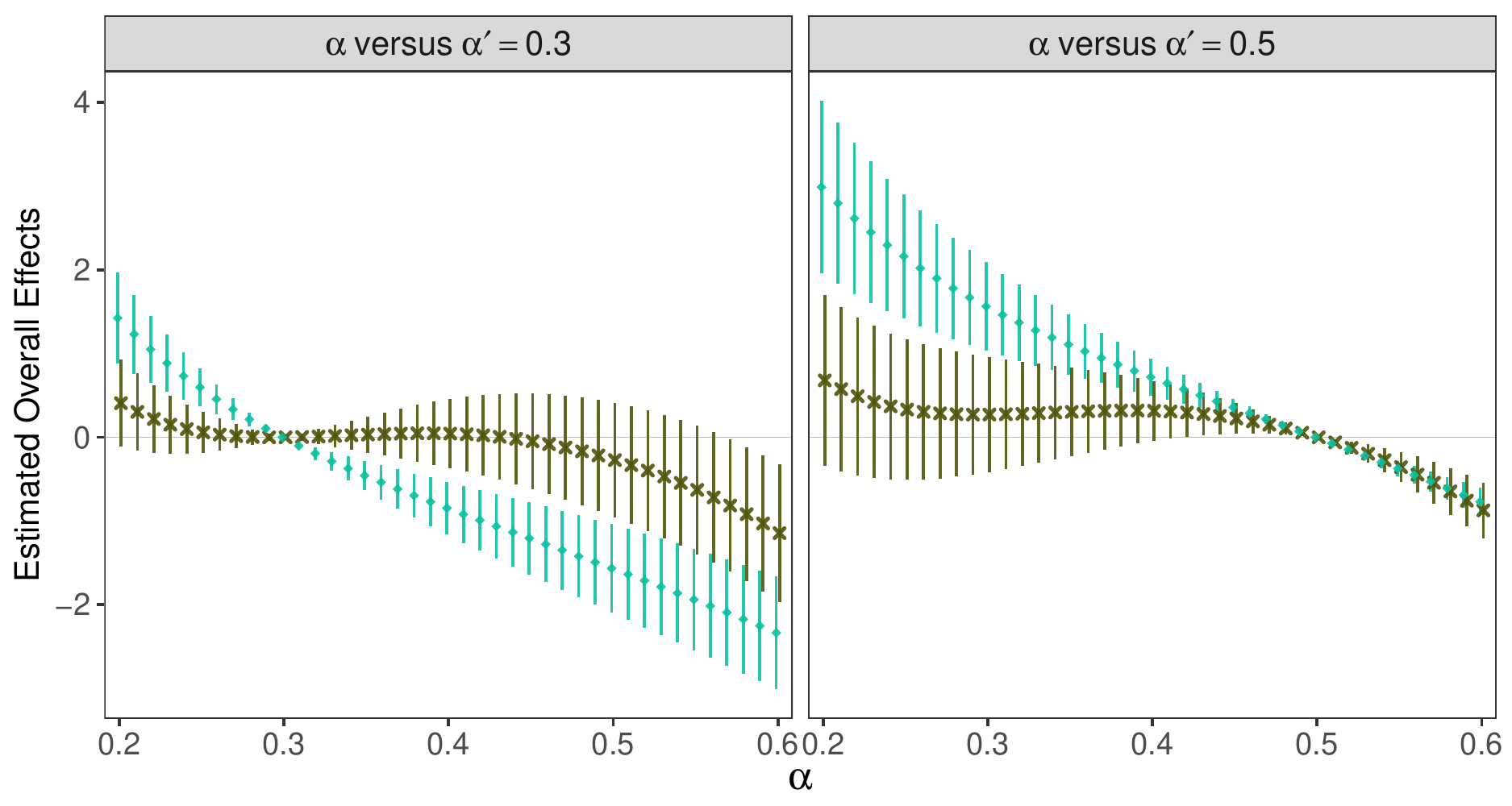}
	\caption{Estimated overall effects from the analysis of the Matlab cholera vaccine trial for selected contrasts. 
		The diamonds and light green lines indicate the point estimates and 95\% CIs from $\widehat{\OVE}(\alpha, \alpha',k=3)$. 
		The $\times$'s and dark brown lines indicate the point estimates and 95\% CIs from $\hOETV(\alpha, \alpha')$, which target the type B estimands from \citet{Tchetgen2012OnCausalInference}. 
		All estimates are multiplied by 1000. 
		This figure appears in color in the electronic version of this article.
		}
	\label{fig:MCVT_CE_plot}
\end{figure}

Overall effect estimates and corresponding 95\% CIs are depicted in Figure~\ref{fig:MCVT_CE_plot}. Negative effects are favorable, corresponding to a reduction in cholera infections. For example, $\widehat{\OVE}(0.45,$ $0.3, k=3)=-1.2$ (95\% CI $-1.6, -0.8$), indicating a significant protective effect of policy $\alpha=0.45$ compared to $\alpha = 0.3$. In particular, we expect 1.2 fewer cases of cholera per 1000 person-years if there is 45\% vaccine coverage compared to 30\% vaccine coverage. 

\begin{figure}
	\centering
	\includegraphics[width=1\linewidth]{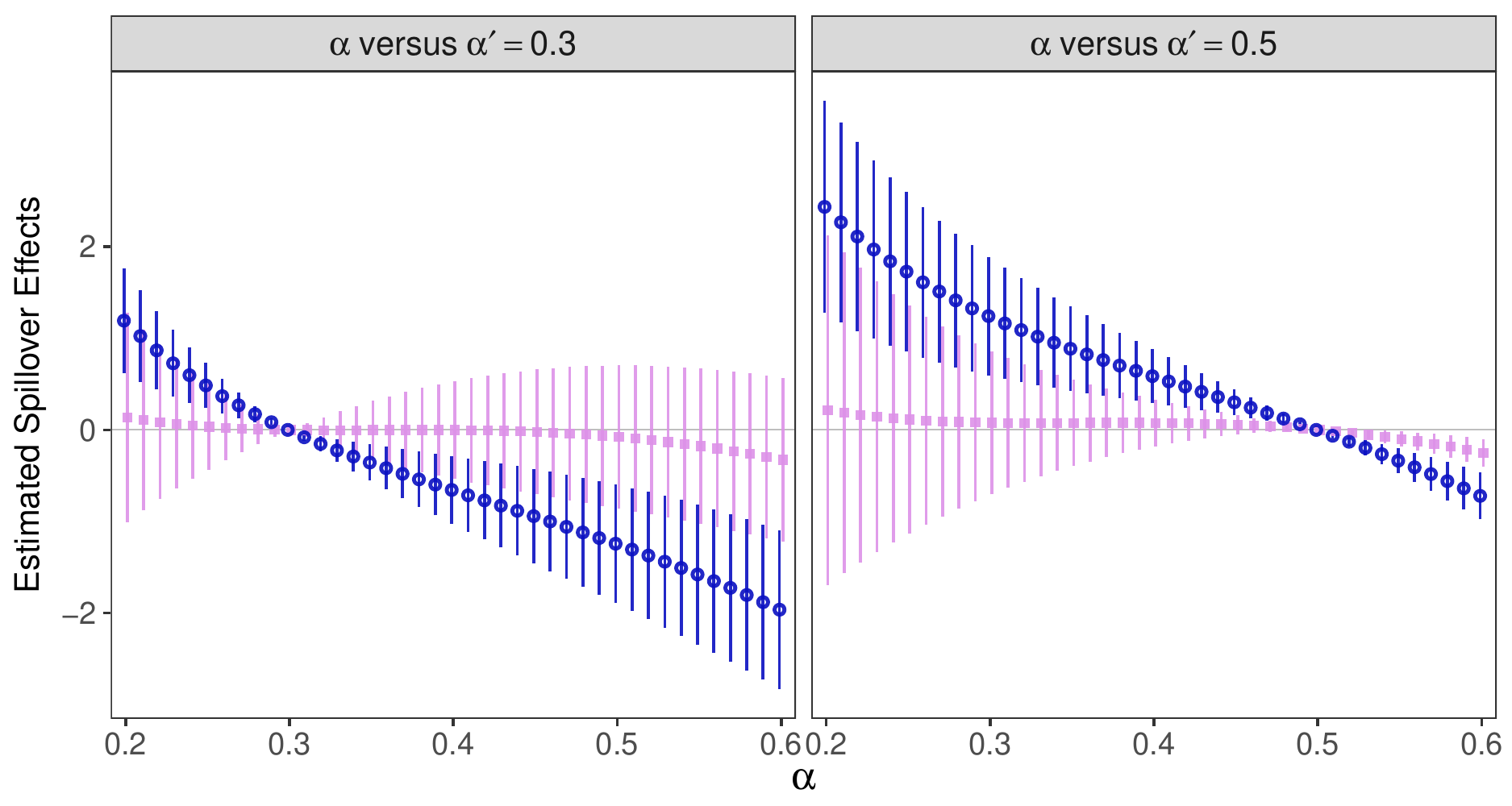} 
	\caption{
		Estimated spillover effects from the analysis of the Matlab cholera vaccine trial for selected contrasts. 
		The circles and dark blue lines indicate the point estimates and 95\% CIs from $\widehat{\SE}_0(\alpha, \alpha', k=3)$. 
		The squares and light pink lines indicate the point estimates and 95\% CIs from $\widehat{\SE}_1(\alpha, \alpha', k=3)$. 
		All estimates are multiplied by 1000. 
		This figure appears in color in the electronic version of this article.
	}
	\label{fig:MCVT_SE_plot}
\end{figure}

Estimated spillover effects are depicted in Figure~\ref{fig:MCVT_SE_plot}. The estimates of $\widehat{\SE}_1(\alpha, \alpha',k=3)$ were approximately zero and the CIs included zero for almost all contrasts shown, indicating mostly negligible spillover effects among treated individuals within clusters.  However, $\widehat{\SE}_0(\alpha, \alpha',k=3)$ was negative for $\alpha > \alpha' $ and positive for $\alpha < \alpha'$, and all of the CIs excluded zero. Thus there is evidence of a protective effect of policies with higher probability of treatment exposure conferred to individuals who did not themselves obtain treatment.

Figures~\ref{fig:MCVT_mus_plot} and~\ref{fig:MCVT_CE_plot} also depict point estimates of the type B estimands and corresponding 95\% CIs, computed using the \texttt{R} package \texttt{inferference} \citep{Saul2017Recipe} based on the same logistic mixed effects propensity score model employed with the proposed estimators. Relative to the estimates of the proposed estimands, the estimates of the type B estimands were smaller with corresponding 95\% CIs that often included zero. For example, $\hOETV(0.2, 0.5)=0.7$ (95\% CI $-0.3, 1.7$), whereas $\widehat{\OVE}(0.2, 0.5,k=3 )=3.0$ (95\% CI $2.0, 4.0$). That is, there is evidence of lower population-level risk of cholera infection due to increased vaccine coverage arising from the policies of interest, but the same is not generally true for the type B policies.

\section{Discussion} \label{sec:Discussion} 
Drawing causal inference from observational data when interference may be present poses several challenges,
including determining an appropriate definition of causal effects. Proposed in this paper are causal estimands for use in observational studies when clustered interference is plausible. The proposed causal effects are contrasts in mean potential outcomes arising from different policies that change the distribution of treatment. IPW estimators were proposed and shown to be consistent and asymptotically Normal under certain identifying assumptions, and empirical sandwich estimators were derived for the asymptotic variance of the estimators. The IPW estimators performed well in finite samples with minimal bias, and the Wald-type confidence intervals attained nominal coverage. These methods were illustrated in an analysis of a large cholera vaccine study, providing evidence that increasing the proportion of individuals vaccinated reduces cholera infections.

The policies discussed here may be more relevant in public health settings such as infectious disease research because within-cluster relationships are incorporated into the proposed estimands. For example, the rank-preserving assumption in Section~\ref{sec:IDable} supposes that the conditional odds ratio of treatment for any two individuals within the same cluster is the same across the factual and counterfactual scenarios. Such orderings are not preserved in the type B estimands.
Although independent exposure of individuals to treatment can be an valuable tool for drawing inference about some causal estimands because potential confounding is mitigated (as in a randomized controlled trial), these scenarios may not be of interest when interference is present. 
\citet{ali2009modeling} discuss that accounting for the ``ecological circumstances'' of infectious diseases can assist in vaccination programs beyond the evidence that can be gained from a controlled trial. Thus, by considering non-independence of treatment selection as well as interference, the proposed estimands may play a role in reducing the burden of infectious disease.

Consistency of the IPW estimators considered in this paper requires the parametric propensity score models be correctly specified. Although non-parametric methods might be employed instead to improve robustness to model mis-specification, such methods may impede identifiability of the target causal estimands without further untestable identifying assumptions. Another drawback of the proposed IPW estimators is that some instances require estimation of a large number of nuisance parameters, which can present computational challenges. Future work may consider reducing the number of nuisance parameters, perhaps through approximating the counterfactual treatment distribution. Considering alternative parameterizations may also allow for relaxing the assumption of no interference between baris to align with developing research on the pathology of cholera \citep{ali_accepted}.

Although this work is motivated by infectious disease research, it is applicable in many other areas in which interference may be present. For example, \citet{Papadogeorgou2017ArXiv} are currently and independently developing similar estimands and methods with motivation from and applications in air pollution epidemiology. By defining causal effects of population-level interventions \citep{westreich2017patients} in the presence of interference, the proposed estimands may be more relevant to investigators and have the potential to impact scientific discovery.

\backmatter

\section*{Software}

An R software package implementing the proposed inverse probability-weighted estimators is provided at \url{https://github.com/BarkleyBG/clusteredinterference}. 

\section*{Acknowledgments}
We would like to thank Wen Wei Loh, Bradley Saul, and Betz Halloran for their helpful comments and advice. This work was partially supported by NIH grants R01 AI085073 and T32 ES007018. 
\bibliographystyle{biom}
\bibliography{ClusteredInterferenceBib}
 
\appendix 

\newpage
\begin{center}
	\textsc{Appendices} \vspace{-0.2 in}
\end{center}
\section*{\hypertarget{sec:EEqn_var}{Appendix A.} Large-sample estimating equation theory} 

The IPW estimators introduced in Section~\ref{sec:Estimators} are shown to be consistent and asymptotically Normal using standard large-sample estimating equation theory or ``M-estimation'' \citep{Stefanski2002Calculus}. Presented for illustration below is a simple example where each cluster has exactly $n$ individuals, and at least one cluster $i\leq M$ is observed to experience treatment $f(A_i) = s$ for each $s=0,1,\dots,n$. \hyperlink{sec:AppB}{Appendix B} provides a more extended example for illustrating details related to estimating the counterfactual probability terms. Let $\omega_{\alpha}= (\omega(0,n,\alpha), \dots, \omega(n-1,n,\alpha))$ be the ordered vector of the possibly unique counterfactual probabilities; the law of total probability implies that ${\omega}(n,n, \alpha ) = 1-\sum_{s=0}^{n-1} {\omega}(s,n,\alpha )$ for the remaining counterfactual probability. Let $\theta_{\alpha }= \left( \beta_0, \beta_1, \sigma,\gamma_{0\alpha} ,\omega_{\alpha}, \mu(\alpha ) \right)$ be the ordered vector of all parameters to estimate. Next, estimating functions corresponding to each element of $\theta_{\alpha }$ are introduced.

Estimating functions for the parameters $\nu = (\beta_0, \beta_1, \sigma)$ in the logistic mixed model are based on the score equations for the model. For $\beta_1$, 
\begin{equation*} 
\psi_{\beta_1}(O_i;\theta_{\alpha })= {\frac{\partial}{\partial {\beta_1}}} \log\big\{\! \Pr(A_i| L_i, N_i )\big\},
\end{equation*}
where $ \Pr(A_i  | L_i, N_i)$ is given in \eqref{eq:LMEM}. Let $\psi_{\nu}  = ( \psi_{\beta_0},\psi_{\beta_1} , \psi_{\sigma} )^\intercal$ be a column vector of estimating functions, with slight abuse of notation in the omission of the functions' inputs. For $\gamma_{0\alpha}$, define the estimating function
\begin{equation*} 
\psi_{\gamma_{0\alpha} }(O_i; \theta_{\alpha }) =
\Bigg\{N_i^{-1}\sum_{j=1}^{N_i} \int \expit({\gamma}_{0\alpha } + {\beta}_{1}L_{ij} + b_i) d\Phi(b_i;{\sigma})\Bigg\} -\alpha.
\end{equation*} 
For each $\omega(s,n,\alpha ) \in \omega_{\alpha}$, define the estimating function
\begin{equation*}\label{MCFPeqn:omega_z} 
\psi_{\omega(s,n,\alpha )}(O_i; \theta_{\alpha }) = \Bigg\{ \sum\limits_{a \in \mathcal{A}(n, s)}
{\Pr}_{\alpha }(A_i = a |L_i , N_i) - \omega(s, n, \alpha ) \Bigg\}I(N_i=n),
\end{equation*} 
and let $\psi_{\omega_{\alpha}} = (\psi_{\omega(0,n,\alpha )} ,\psi_{\omega(1,n,\alpha )} ,\dots, \psi_{\omega(n-1, n, \alpha )} )^\intercal$. For the target estimand, define
\begin{equation*}\label{MCFPeqn:mu}
\psi_{\mu(\alpha )}(O_i; \theta_{\alpha }) = 
\frac{\overline{Y}_i \omega(A_i, N_i, \alpha ) }{{\Pr}(A_i|L_i, N_i)}  - \mu(\alpha ),
\end{equation*}
where $\omega(A_i, N_i, \alpha ) = \binom{N_i}{ f(A_i)}^{-1}\omega(f(A_i), N_i, \alpha )$ and where ${\Pr}(A_i |L_i , N_i)$ is the propensity score for the cluster as in \eqref{eq:LMEM}. 

Let $\psi_{\theta_{\alpha }}  = (\psi_{\nu} , \psi_{\gamma_{0\alpha} } , \psi_{\omega_{\alpha}} , \psi_{\mu(\alpha )} )^\intercal$, and let $q = |\theta_{\alpha }|$ be the number of parameters to estimate. The estimator $\hat{\theta}_{\alpha }$ can be expressed as a solution to the following system of ``stacked'' estimating equations: 
\begin{equation*} \label{MCFPeqn:solvepsialpha2}
\sum\limits_{i=1}^M {\psi_{\theta_{\alpha }}}(O_i; \theta_{\alpha })=\sum\limits_{i=1}^M
\left[\begin{array}{r}
\psi_{\nu} (O_i ; \theta_{\alpha })\\
\psi_{\gamma_{0\alpha} }(O_i; \theta_{\alpha }) \\ 
\psi_{\omega_{\alpha}}(O_i; \theta_{\alpha }) \\
\psi_{\mu(\alpha )}(O_i; \theta_{\alpha })\\
\end{array}\right] =0_{q\times 1}.
\end{equation*}

To show that $\mu(\alpha)$ is the solution to $\int \psi_{\mu(\alpha)}(O | \theta_{\alpha}) dF_O(O)=0$, write
\begin{align*}
\int \psi_{\mu(\alpha)}(O_i | \theta_{\alpha}) dF_O(O_i) 
&= \Eval\Bigg\{\frac{\overline{Y}_i \omega(A_i, N_i, \alpha ) }{{\Pr}(A_i=a_i|L_i, N_i)}  - \mu(\alpha )\Bigg\}
\end{align*}
where the expected value is taken over the joint distribution of observable random variables, i.e., $(N_i,L_i,A_i,\{Y_i(a)\}_{a\in \mathcal{A}(N_i)})$. Then, 
\begin{align*}
 \Eval\Bigg\{\frac{\overline{Y}_i \omega(A_i, N_i, \alpha ) }{{\Pr}(A_i=a_i|L_i, N_i)} \Bigg\} &= \Eval\Bigg\{\sum_{a \in \mathcal{A}(N_i)}\frac{\overline{Y}_i(a)\omega(a, N_i, \alpha ) }{{\Pr}(A_i=a|L_i, N_i)} I(A_i=a) \Bigg\}\\ 
&=  \Eval_{L_i, N_i}\Bigg[ \sum_{a \in \mathcal{A}(N_i)} \bigg\{  \Eval_{A_i, \{Y_i(a)\} | L_i, N_i}\big(\overline{Y}_i(a)\omega(a, N_i, \alpha )\big)\, \times  \\ 
& \hspace{1.56 in} \Eval_{A_i, \{Y_i(a)\} | L_i, N_i}\left( \frac{I(A_i=a)}{{\Pr}(A_i=a|L_i, N_i)}  \right) \bigg\} \Bigg]\\ 
&=  \Eval\Bigg\{\sum_{a \in \mathcal{A}(N_i)}{\overline{Y}_i(a)\omega(a, N_i, \alpha ) }\Bigg\},
\end{align*} 
which equals $\mu(\alpha)$ by definition, and so $\mu(\alpha)$ is the zero of $\int \psi_{\mu(\alpha)}(O_i | \theta_{\alpha}) dF_O(O_i)$. Since $\psi_{\nu}$ are simply the score equations, $\int \psi_{\nu}(O_i | \theta_{\alpha}) dF_O(O_i)=0$. Note that the right side of \eqref{eq:alpha_ID} equals $\alpha + \int\psi_{\gamma_{0\alpha}}(O_i | \theta_{\alpha}) dF_O(O_i)$, and so $\gamma_{0\alpha}$ is the zero of $\int \psi_{\gamma_{0\alpha}}(O_i | \theta_{\alpha}) dF_O(O_i)$. Finally, $\int \psi_{\omega(s,n,\alpha )}(O_i; \theta_{\alpha }) dF_O(O_i)=0$ follows from $\omega(a,n,\alpha ) = \Eval_{L}\{{\Pr}_{\alpha }(A = a | L , N=n)\}$.

From \citet{Stefanski2002Calculus}, $\hat{\theta}_{\alpha } \xrightarrow{p} \theta_{\alpha }$ and $\sqrt{M}(\hat{\theta}_{\alpha } - \theta_{\alpha }) \xrightarrow{d} N({0}, {\Sigma}_{\alpha })$, where ${\Sigma}_{\alpha }= U_{\alpha }^{-1}W_{\alpha }(U_{\alpha }^{-1})^{\intercal}$ for $U_{\alpha }=\Eval\{-\dot{{\psi}}_{\theta_{\alpha }}(O_i; {\theta}_{\alpha })\}$ and $W_{\alpha }=\Eval\{{\psi_{\theta_{\alpha }}}(O_i; {\theta}_{\alpha })^{\otimes 2}\}$. Consistent estimators for $U_{\alpha }$ and $W_{\alpha }$ are $ \widehat{U}_{\alpha }=M^{-1}\sum_{i=1}^M  \{  -\dot{{\psi}}_{\theta_{\alpha }}(O_i; {\theta}_{\alpha })  |_{{\theta}_{\alpha } = \hat{{\theta}}_{\alpha }}  \} $ and $ \widehat{W}_{\alpha } =M^{-1}\sum_{i=1}^M  \{  \psi_{\theta_{\alpha }}(O_i; \hat{\theta}_{\alpha })^{\otimes 2} \}. $ The empirical sandwich variance estimator $\widehat{\Sigma}_{\alpha } =\widehat{U}_{\alpha }^{-1}\widehat{W}_{\alpha }(\widehat{U}_{\alpha }^{-1})^{\intercal}$ is consistent for $\Sigma_{\alpha }$, and so $\widehat{\Var}(\widehat{\mu}(\alpha)) = M^{-1}[\widehat{\Sigma}_{\alpha }]_{[q,q]}$ approximates the variance of $\widehat{\mu}(\alpha)$ for large $M$, where $ [\widehat{\Sigma}_{\alpha } ]_{[q,q]}$ is the bottom-right element of $\widehat{\Sigma}_{\alpha }$. 

An analogous approach is described for $\widehat{\OVE}(\alpha , \alpha' ,k)$, where it is now necessary to estimate $\gamma_{0\alpha' }$ and $\omega_{\alpha'}$ as well. Let 
$\theta_{\alpha , \alpha' }= \left( \nu,\gamma_{0\alpha} ,\gamma_{0\alpha' },\omega_{\alpha},\omega_{\alpha'}, \OVE(\alpha ,\alpha' ) \right)$ be the ordered vector of all parameters to estimate. 
For each $\omega(s,n,\alpha ) \in \omega_{\alpha}$, define the estimating function
\begin{equation*}
\psi_{k,\omega(s,n,\alpha)  }   (O_i;\theta_{\alpha , \alpha' }) =
\left\{ {k^{-1}_{s,n}{\binom{n}{ s}}}\sum\limits_{a \in \mathcal{A}(n,s,k)} 
{\Pr}_{\alpha }(A_i = a|L_i , N_i) - \omega(s, n, \alpha ) \right\} I(N_i=n),
\end{equation*}
and let $\psi_{k,\omega_{\alpha}} = (\psi_{k,\omega(0,n,\alpha )} ,\psi_{k,\omega(1,n,\alpha )} ,\dots, \psi_{k,\omega(n-1, n, \alpha )} )^\intercal$. For the target estimand, define
\begin{equation*} 
\psi_{\OVE(\alpha, \alpha' )}(O_i; \theta_{\alpha, \alpha'}) = 
\frac{\overline{Y}_i \bigl\{\omega(A_i, N_i, \alpha )-\omega(A_i, N_i, \alpha' )\bigr\} }{{\Pr}(A_i|L_i, N_i)} - \OVE(\alpha, \alpha' ).
\end{equation*}
It is easily shown that $\int \psi_{\OVE(\alpha, \alpha' )}(O_i; \theta_{\alpha, \alpha'})  dF_O(O_i)=0$ using a proof analogous to the one for $ \psi_{\mu(\alpha)}$ presented above. Similarly, $\int \psi_{k,\omega(s,n,\alpha)  }(O_i;\theta_{\alpha , \alpha' }) dF_O(O_i)=0$ follows directly from $\int \psi_{\omega(s,n,\alpha )}(O_i; \theta_{\alpha }) dF_O(O_i)=0$. Finally, let $\psi_{k,\theta_{\alpha ,\alpha'}} = (\psi_{\nu}, \psi_{\gamma_{0\alpha} }, \psi_{\gamma_{0\alpha'} },  \psi_{k,\omega_{\alpha}},$ $\psi_{k,\omega_{\alpha'}}, \psi_{\OVE(\alpha , \alpha' )})^{\intercal}$. Then $\hat{\theta}_{\alpha ,\alpha'}$ solves $\sum_{i=1}^M \psi_{k,\theta_{\alpha ,\alpha'}}(O_i ; \theta_{\alpha ,\alpha'}) =0_{|\theta_{\alpha ,\alpha'}|\times 1}$ and the above results follow. 

Notably, the difference in $\widehat{\OVE}(\alpha , \alpha' )$ and $\widehat{\OVE}(\alpha , \alpha' ,k)$ lies in the estimating functions used for the counterfactual probabilities, i.e., $\psi_{\omega_{\alpha}}$ and $\psi_{k,\omega_{\alpha}}$, respectively. For example, when $ \lfloor n/2 \rfloor \leq k $ then $\mathcal{A}(n,s)=\mathcal{A}(n,s,k)$ for all $s$ and $\psi_{\omega_{\alpha}}$ is equivalent to $\psi_{k,\omega_{\alpha}}$. As mentioned in the main paper, an extension of this method is to use different values of $k$ for distinct estimating equations. For example, one could estimate $ \omega(s,n,\alpha)$  with $\psi_{k,\omega(s,n,\alpha) }$ and $ \omega(s',n',\alpha )$ with $\psi_{k',\omega(s',n',\alpha )}$, where  $ \omega(s,n,\alpha) \neq \omega(s',n',\alpha )$ and $k\neq k'$, and the above results would still apply.

\section*{\hypertarget{sec:AppB}{Appendix B.} Estimating the counterfactual probabilities} 
\subsection*{\small{\textsc{\hypertarget{sec:AppB1}{B.1}. Choice of estimator}}}

Some considerations for estimating the counterfactual probabilities $\omega(a,n,\alpha)$ are described below. 
All assumptions for identification discussed in the main paper Section~\ref{sec:IDable} are also made here; in particular that the ordering of individuals within clusters to be uninformative.

Let there be a random sample of $i=1,\dots, M$ clusters, and as in the main paper denote by $O_i=\{N_i, L_i, A_i, Y_i \}$ the observed values of the random variables for cluster $i$. 
As described in Section~\ref{sec:Estimators} of the main paper, $\widehat{\Pr}_{\alpha }(A_i=a|L_i, N_i)$ is calculated by substituting the estimates $(\hat{\gamma}_{0\alpha } , \hat{\beta}_1, \hat{\sigma})$ into the counterfactual cluster propensity score, $ {\Pr}_{\alpha }(A_i=a|L_i, N_i)$.
An estimator for the counterfactual probabilities is
\[
\widetilde{\omega}(a,n,\alpha ) =  \left\{\sum_{i=1}^M I(N_i=n)\right\}^{-1} 
\sum_{i=1}^M \widehat{\Pr}_{\alpha }(A_i=a|L_i, N_i), 
\] 
which is not employed in the main paper for the reasons described below.

Define $f(a)=\sum_{j=1}^n a_j$ to be the sum of the binary entries of $a\in \mathcal{A}(n)$. 
Letting $a,a'\in \mathcal{A}(n)$ be two vectors such that $f(a)=f(a')$, the assumed irrelevance of within-cluster ordering of individuals supposes that $\omega(a,n,\alpha )=\omega(a', n, \alpha )$. However, in any finite sample it is likely that $\widetilde{\omega}(a,n,\alpha ) \neq \widetilde{\omega}(a',n,\alpha)$, which is an undesirable property of the above estimator. Thus, the estimator $\widetilde{\omega}(a,n,\alpha)$ is not pursued further here nor in the main paper.

The method presented in Section~\ref{sec:Estimators} of the main paper is discussed in further detail here.
Under this assumption that the ordering of individuals within clusters to be uninformative, the counterfactual probabilities for clusters of size $n$ and for a policy $\alpha$ take on a maximum of $n+1$ unique values, rather than $2^n=|\mathcal{A}(n)|$. These counterfactual probabilities then arise from the strata of $\mathcal{A}(n,s) =\{a \in \mathcal{A}(n) | f(a)=s\}  $ for $s=0,1,\dots, n$, such that:
\begin{equation*}\label{def:omega_z}
\omega(s,n, \alpha ) = \sum\limits_{a \in \mathcal{A}(n,s)} \omega(a,n, \alpha ). 
\end{equation*} 
Thus for each $a\in \mathcal{A}(n)$ the counterfactual probabilities can be written as $\omega(a, n, \alpha ) = \binom{n}{f(a)}^{-1}\omega(f(a), n, \alpha )$, and estimated by
\begin{equation*}\label{est:omega_a}
\widehat{\omega}(a,n,\alpha ) = \binom{n}{f(a)}^{-1}\widehat{\omega}(f(a), n, \alpha ),
\end{equation*}
where $\widehat{\omega}(f(a), n, \alpha )$ is obtained by
\begin{equation*}\label{est:omega_z}
\widehat{\omega}(f(a),n, \alpha ) = \left\{\sum\limits_{i=1}^M I(N_i=n)\right\}^{-1} \sum\limits_{a \in \mathcal{A}(n,f(a))} 
\sum\limits_{i=1}^M  \widehat{\Pr}_{\alpha }(A_i=a|L_i , N_i ).
\end{equation*}

\subsection*{\small{\textsc{B.2. An example}}} 
It is assumed that there is an upper bound to the cluster size, so the number of counterfactual probabilities to estimate is bounded. The number of possible parameters to estimate is dependent on the sample of data; it may not be necessary to estimate all possible combinations of the counterfactual probabilities.
An example is presented here for illustrating estimating the counterfactual probabilities by the strata of $\mathcal{A}(n,s)$. Let there be $i=1,\dots, M$ clusters in this sample, and assume $M=M_{n}+M_{n'}$ where each of the $M_{n}$ clusters contains exactly $n$ individuals, and each of the $M_{n'}$ clusters contains exactly $n'$ individuals. For ease of notation and without loss of generality, order the clusters so that each of the clusters $i=1, \dots, M_{n}$ has $n$ individuals, and each of the clusters $i=(M_{n}+1), \dots, M$ has $n'$ individuals. 

Let $\mathcal{P}(n)\subsetneq \{0, \dots, n\}$ be a proper subset, and assume that at least one cluster $i=1,\dots,M_n$ in the now-ordered sample is observed to have treatment $f(A_i)=s$ for $s \in \mathcal{P}(n)$. Further assume that none of these clusters are observed to be exposed to treatment $A_i$ such that $f(A_i) \notin \mathcal{P}(n)$. In this case, only the values of $\omega(s,n, \alpha )$ for $s \in \mathcal{P}(n)$ must be estimated. Order the elements $s \in \mathcal{P}(n)$ to be $(s_{1,n}, \dots, s_{p(n),n})$ where  $p(n) = |\mathcal{P}(n)|$ and $s_{j,n} < s_{j+1,n}$ for any $j< p(n)$.

Next, for each $s=0,1,\dots,n'$, assume that at least one cluster $M_{n}< i \leq M$ is observed to have treatment $f(A_i) = s$. Direct estimation of $\omega(s,n', \alpha )$ is only necessary for $n'$ of these counterfactual probabilities; e.g., estimate $\widehat{\omega}(s,n',\alpha )$ directly for $s=0,1,\dots, (n'-1)$, and calculate $\widehat{\omega}(n',n', \alpha ) = 1-\sum_{s=0}^{n'-1} \widehat{\omega}(s,n',\alpha )$. 

Now define the ordered set of all counterfactual probabilities to be estimated to be $\omega_{\alpha}= \left( \omega(s_{1,n}, n, \alpha ), \dots, \omega(s_{p(n),n}, n, \alpha ) ,\omega(0,n',\alpha ),\omega(1,n', \alpha ), \dots, \omega(n'-1, n', \alpha ) \right)$. When treatment is not randomized, it is necessary to estimate all of the parameters in the vector $\theta_{\alpha }= \left( \beta_0, \beta_1, \sigma,\gamma_{0\alpha} ,\omega_{\alpha},  \mu(\alpha ) \right)$ in order to obtain an estimate of $\widehat{\mu}(\alpha)$. Estimation and inference can be carried out by standard estimating equation theory \citep{Stefanski2002Calculus} as shown in \hyperlink{sec:EEqn_var}{Appendix A}.

\section*{\hypertarget{sec:AppC}{Appendix C.} Determining true values of estimands for simulation study}

The true values of target estimands and nuisance causal parameters were determined empirically, as explained below. Recall the steps for generating a sample of data described in the main paper: for each cluster $i$, step~\ref{step1} was to generate the number of individuals in the group $N_i$, step~\ref{step2} was to simulate covariates $L_i$, step~\ref{step3} was to generate an observed treatment vector $A_i$, and step~\ref{step4} was to generate the observed outcome $Y_i$.

\subsection*{\small{\textsc{\hypertarget{sec:AppC1}{C.1}. Determining counterfactual intercepts}}}
To determine $\gamma_{0\alpha}$ for $\alpha  \in \{0.40, 0.50, 0.55, 0.75\}$, a grid of $W$-many potential values $\gamma_{1}^* < \gamma_{2}^* < \dots < \gamma_{w}^* < \dots < \gamma_{W}^*$ was proposed. For each $w= 1, \dots, W$, the following steps were carried out:
\begin{enumerate}
	\item Steps~\ref{step1} and~\ref{step2} were repeated for $m_1=10^7$ clusters.
	\item Treatment vectors were generated under policy $\alpha$ for $m_1$ clusters by replacing $\beta_0$ in step~\ref{step3} with $\gamma_{w}^*$. That is, $A_{ij,w}$ for each individual $j$ in cluster $i$ was simulated from a Bernoulli distribution with probability $\expit(\gamma_{w}^* -0.015L_{1ij} -0.025L_{2ij}+b_i)$ where $b_i\sim N(0, 0.75)$.
	\item The probability of obtaining treatment was assumed to equal the proportion of individuals in the dataset obtaining treatment, $p_w=\sum_{i=1}^{m_1} \sum_{j=1}^{N_i} I(A_{ij,w} =1) / (\sum_{i=1}^{m_1} N_i)$.
\end{enumerate}
For each $\alpha$, $\gamma_{0\alpha} $ was determined to be the average of the $\gamma_{w}^*$ that produced probabilities $p_w$ closest to $\alpha$, i.e., $\gamma_{0\alpha} = \mean(\gamma_{w_{l}}^*, \gamma_{w_{u}}^*)$ where $ w_{l} = \argmax\limits_{\{w| p_w < \alpha \}} ( \alpha -p_w )$ and $ w_{u}  = \argmax\limits_{\{w| p_w > \alpha \}} ( p_w-\alpha  )$.

\subsection*{\small{\textsc{\hyperlink{sec:AppC2}{C.2.} Determining counterfactual probabilities}}}
For each $\alpha $, $\omega(a,n,\alpha )$ was determined empirically from values of $\gamma_{0\alpha} $ determined as above. For each $n=8,22,40$ and each $\alpha$ the following steps were carried out:
\begin{enumerate}
	\item Step~\ref{step2} was repeated for $m_2=10^8$ clusters of fixed size $n$.
	\item Treatment vectors were generated under policy $\alpha$ for $m_2$ clusters by replacing $\beta_0$ in step~\ref{step3} with the value $\gamma_{0\alpha}$ determined in \hyperlink{sec:AppC1}{Appendix C.1}. That is, $A_{ij,\alpha}$ for each individual $j$ in cluster $i$ was simulated from a Bernoulli distribution with probability $\expit(\gamma_{0\alpha} -0.015L_{1ij} -0.025L_{2ij}+b_i)$ where $b_i\sim N(0, 0.75)$.
	\item The counterfactual probabilities was defined as $\omega(a,n,\alpha )=\binom{n}{s}^{-1}\omega(s,n,\alpha )$ for each $s=0, 1, \dots, n$ where $\omega(s,n,\alpha ) = m_2^{-1} \sum_{i=1}^{m_2} I\left(\sum_{j=1}^{n} A_{ij,\alpha }  = s\right)\!\!.$  
\end{enumerate}	

\subsection*{\small{\textsc{\hyperlink{sec:AppC3}{C.3}. Simulating potential outcomes}}}
For each $n=8,22,40$ and each $s=0, 1, \dots, n$, let $a_{n,s}$ be the vector with $s$ 1's followed by $(n-s)$ 0's. For each $n=8,22,40$, the following steps were carried out:

\begin{enumerate}
	\item Step~\ref{step2} was repeated for $m_3=10^8$ clusters of fixed size $n$.
	\item For each $s=0,1,\dots, n$, 
	\begin{enumerate}
		\item Individual potential outcomes $Y_{ij}(a_{n,s})$ were generated via the causal model analogous to the regression model specified in step~\ref{step4} for all individuals $j$ in each cluster $i$. That is, $Y_{ij}(a_{n,s})$ was 
		simulated from a Bernoulli distribution with mean 
		$\Pr(Y_{ij}(a)=1 | L_{ij}) = \expit(0.1 - 0.05L_{1ij} + 0.5L_{2ij} -0.5a_{j} + 0.2g(a_{-j}) -0.25a_{j}g(a_{-j})  )$,
		where $g(a_{-j}) = (N_i-1)^{-1}\sum_{j'\neq j} a_{j'}$.
		\item Then, $\overline{Y}_i(a_{n,s})$, $\overline{Y}_{0,i}(a_{n,s})$ and $\overline{Y}_{1,i}(a_{n,s})$ were computed for each cluster $i$ according to their definitions presented in Section~\ref{sec:Estimands} of the main paper.
		\item Finally, define $\overline{\overline{Y}(a_{n,s})}=m_3^{-1} \sum_{i=1}^{m_3}\overline{Y}_i(a_{n,s})$ to be the average potential outcomes for all clusters when exposed to treatment $a_{n,s}$. For $t=0,1$ define $\overline{\overline{Y}_t(a_{n,s})}=m_3^{-1} \sum_{i=1}^{m_3}\overline{Y}_{t,i}(a_{n,s})$.
	\end{enumerate}
	
\end{enumerate}

\subsection*{\small{\textsc{C.4. Determining values of target estimands}}}
The values produced in Appendices \hyperlink{sec:AppC2}{C.2} and \hyperlink{sec:AppC3}{C.3} were combined to determine the values of the target estimands. That is, 
\[\mu(\alpha ) = \sum\limits_{n \in \{8,22,40\}} \Biggl\{\sum\limits_{s=0}^n \biggl(\overline{\overline{Y}(a_{n,s})}\omega(s,n,\alpha )\biggr) \Pr(N_i=n)\Biggr\},\]
and $\OVE(\alpha ,\alpha' ) =\mu(\alpha )- \mu(\alpha' )$. For $t=0,1$,
$\SE_t(\alpha ,\alpha' ) =\mu_t(\alpha )- \mu_t(\alpha' )$, where $\mu_t(\alpha ) = \sum_{n \in \{8,22,40\}} \{\sum_{s=0}^n (\overline{\overline{Y}_t(a_{n,s})}\omega(s,n,\alpha )) \Pr(N_i=n)\}$.

\subsection*{\small{\textsc{C.5. Empirical comparison of proposed and existing methods}}}

\begin{figure} 
	\centering 
	\includegraphics[width=0.99\linewidth]{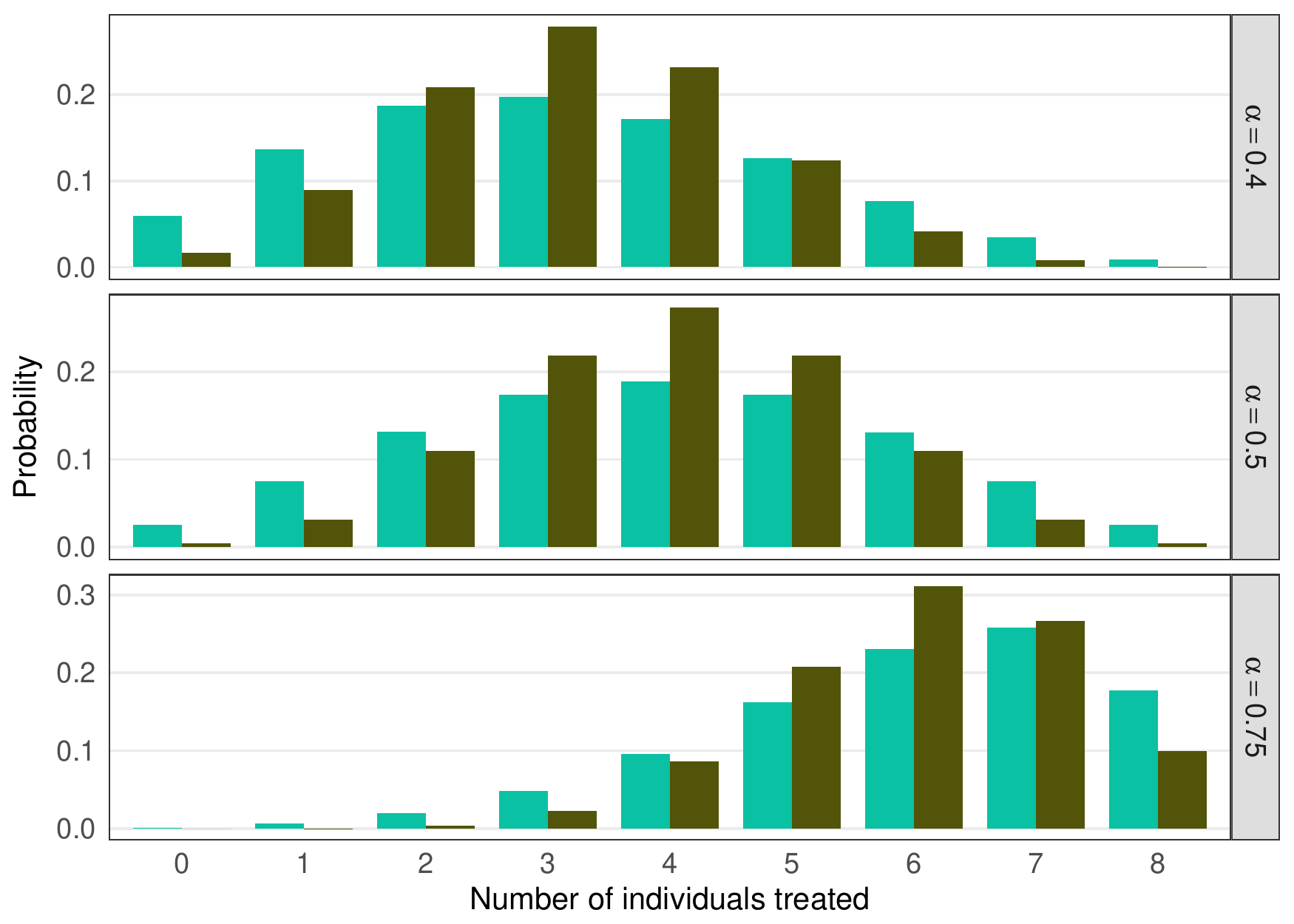} 
	\caption[Comparison of proposed and existing counterfactual treatment distributions]{An empirical comparison of the counterfactual probabilities for the proposed estimands and the type B estimands. The light green bars indicate $\omega(s, n, \alpha )$ and the dark brown bars indicate $\omegaTV(s,n,\alpha )$ for the type B policies from \citet{Tchetgen2012OnCausalInference} for $s\in \{0,1,\dots, 8\}$, $n=8$, and $\alpha  \in \{0.4, 0.5, 0.75\} $ for the data generating process in the simulation study described above and in the main paper.}
	\label{fig:omega_pi_compare2017-08-07}
\end{figure}

Numerical differences in $\omega(a,n,\alpha)$ and $\omegaTV(a,n,\alpha)$ for the type B policies from \citet{Tchetgen2012OnCausalInference} are dependent on the context and data generating process. Figure \ref{fig:omega_pi_compare2017-08-07} depicts the values of $\omega(s,n=8, \alpha )$ determined in \hyperlink{sec:AppC2}{Appendix C.2} and the values of $\omegaTV(s,n=8, \alpha ) = \sum_{a \in \mathcal{A}(n,s)} \omegaTV(a,n=8, \alpha )$. This figures illustrates the inequality $\omega(s,8, \alpha )\neq \omegaTV(s,8, \alpha )$ for all pairs of $s$ and $\alpha$ for the data generating process described above.
The values of $\omega(s,n,\alpha)$ and $\omegaTV(s,n,\alpha)$ are particularly different when $s$ is close to 0 or to $n$. For example, $\omega(0,8,0.40)=0.059$ and $\omegaTV(0,8,0.40)=0.017$, and so for the data generating process in this simulation study the proposed estimands confer $0.059/0.017=3.5$ times more weight to this category than the type B estimands.

\label{lastpage}
\end{document}